\documentclass[12pt]{iopart}

\usepackage{mathrsfs}

\usepackage{graphicx}
\usepackage{psfrag}
\usepackage{color}
\usepackage{braket}
\usepackage{enumerate}
\usepackage{iopams}
\usepackage{color}

\expandafter\let\csname equation*\endcsname\relax

\expandafter\let\csname endequation*\endcsname\relax

\usepackage{amsmath}

\usepackage{cite}

\newcommand{\Bs}{\setminus}
\newcommand{\To}{\rightarrow}
\newcommand{\IT}{\textit}
\newcommand{\be}{\begin{equation}}
\newcommand{\ee}{\end{equation}}
\newcommand{\bea}{\begin{eqnarray}}
\newcommand{\eea}{\end{eqnarray}}
\newcommand{\qqq}{\end{document}}
\newcommand{\bspl}{\begin{split}}
\newcommand{\espl}{\end{split}}

\newcommand{\bV}{\begin{pmatrix}}
\newcommand{\eV}{\end{pmatrix}}
\newcommand{\de}{\partial}

\newcommand{\DE}{\overset{\centerdot}}
 \newcommand{\DEEE}{\overset{\centerdot\centerdot\centerdot}}
  \newcommand{\DEE}{\overset{\centerdot\centerdot}}

\newcommand{\R}{\mathbb{R}}
\newcommand{\Z}{\mathbb{Z}}
\newcommand{\Ha}{\mathcal{H}}
 \newcommand{\K}{\mathcal{K}}
  
    \newcommand{\E}{\mathcal{E}}
    \newcommand{\sgn}{\text{sgn}}
    \newcommand{\Zz}{\frac{\overset{\centerdot}{\zeta}_0\zeta_0}{\omega_0}}
\newcommand{\Nn}{\frac{\overset{\centerdot}{\eta}_0}{2\omega_0}}    
     \newcommand{\EL}{\mathcal{E}\mathcal{L}}
 \newcommand{\besub}{\begin{subequations}}
\newcommand{\eesub}{\end{subequations}}

\newcommand{\GS}{\text{\begin{small}{GS}\end{small}}}

\begin{document}
\title{Exact dynamics of a one dimensional Bose gas in a periodic time-dependent harmonic trap  }
\author{Stefano Scopa $^{a}$, J\'eremie Unterberger $^{b}$ and Dragi Karevski $^{a}$}
\address{$^a$ Laboratoire de Physique et Chimie Th\'eoriques, UMR CNRS 7019, Universit\'e de Lorraine BP 239 F-54506 Vandoeuvre-l\'es-Nancy Cedex, France\\
$^b$ Institut Elie Cartan, UMR CNRS 7502, Universit\'e de Lorraine BP 239 F-54506 Vandoeuvre-l\`es-Nancy Cedex, France 
}
\ead{\mailto{\\ stefano.scopa@univ-lorraine.fr \\ jeremie.unteberger@univ-lorraine.fr \\ dragi.karevski@univ-lorraine.fr}}
\begin{abstract}
We study the unitary dynamics of a one-dimensional gas of hard-core bosons trapped into a harmonic potential which varies periodically in time with frequency $\omega(t)$. Such periodic systems can be classified into orbits of different monodromies corresponding to two different physical situations, namely the case in which the bosonic cloud remains stable during the time-evolution and the case where it turns out to be unstable.  In the present work we derive in the large particle number limit exact results for the stroboscopic evolution of the energy and particle densities in both physical situations.
\end{abstract}
\section{Introduction}
Understanding the dynamics of interacting quantum many body systems is a fundamental issue which has motivated a lot of theoretical and experimental investigations these last years. In particular the focus was especially put on the out of equilibrium dynamical properties  which are still very poorly understood compared to their equilibrium counterpart. 
Nevertheless, significant progress has been achieved for certain integrable one-dimensional systems for which the unitary relaxation after a sudden quench has been fully understood in terms of Generalized Gibbs States  \cite{term1,term2,term3,term4,term5,term6,term7,term8} and the emergence of generalized hydrodynamics \cite{doyon1,bertini1}. 
For space-time inhomogeneous systems, built up by slow space-time variation of the local coupling constants, general scaling theories have been proposed \cite{nonlin,PKT07,gradient,critical,VIC3,VIC,gritsev1} which lead to generalized predictions for Kibble-Zurek topological defects proliferation, see \cite{Dziarmaga,polkovnikov} for recent reviews. In that direction one may mention the recent proposals for shortcuts to the adiabatic evolution relying on the use of auxiliary fields \cite{noAD1,noAD2,noAD3,noAD4,noAD5,noAD6,noAD7}, ultimately preventing the generation of Kibble-Zurek defects.   

 Most of these studies have been motivated by the impressive experiments performed with ultracold atoms loaded onto quasi-one-dimensional optical traps \cite{13,14,15,16,17,18,19}. These systems are well described at the theoretical level by a Bose-Hubbard model in the hard-core limit,  that is a Tonks-Girardeau gas of bosons in a lattice.
 The main difficulty in predicting the out of equilibrium dynamics of such bosonic gases after a typical quantum quench-like experiment  relies on the fact that one has to take into account a huge sum of coherent states and not merely the evolution of the instantaneous ground state and few low-lying excited states. 
 However, this kind of problems can be approached with the help of time-dependent perturbative expansions \cite{nonlin,VIC,Scopa} for slowly varying systems, or by the construction of dynamical invariants and their relative eigenstates, in terms of which the dynamics can be exactly determined \cite{Scopa,MING,Kagan}. In our recent work \cite{Scopa} we applied both perturbative techniques and the use of dynamical invariants to the investigation of the dynamics of a Tonks Girardeau gas \cite{TG1,TG2,TG3} during the release of an harmonic trap potential. One may also mention the recent adaptation of generalized hydrodynamics for inhomogeneous systems \cite{dubail} reproducing the observations made in the famous quantum Newton's cradle experiment \cite{13}.  

For periodically driven quantum gases, one may use Floquet's type approaches  \cite{per1,per2,per3,per4,per5,per6,per7,per8,per9,per10} in order to investigate the dynamics, typically putting a special focus on the stroboscopic evolution while leaving aside the so called micromotion.  
The aim of this work is instead to use dynamical invariants for studying such periodically driven systems. 
To be more specific, we consider a low-density bosonic gas of impenetrable bosons in one dimension, well-described by the Tonks Girardeau model, loaded on a harmonic trap which has a generic $2\pi$-periodic dependence in time. 

By noticing that the single particle hamiltonian for this system can be reduced (in the thermodynamic limit) to a time-dependent harmonic oscillator with Schr\"odinger operator $\mathcal{S}=i\de_t-\frac{1}{2}(-\de_x^2+\omega^2(t) x^2)$ and periodic frequency  $\omega(t+2\pi)=\omega(t)$, we can classify the system into classes associated to different monodromies \cite{JER,RogUnt}. 
These classes, depending on the function $\omega(t)$, correspond physically to  stable and unstable phases during time evolution. 
One of the main advantages of this classification lies in the fact that Schr\"odinger operators belonging to the same orbit within a specific class are connected by a simple, orientation preserving, time reparametrization ($t\mapsto t'=\varphi(t)$, $\varphi\in\text{Diff}_+(\R/2\pi\Z)$) and one needs to know only a single representative to understand the full orbit. 

The essential remark here is (as noted initially
in \cite{JER}) that time-reparametrization orbits of the space
of periodic time-dependent harmonic oscillators is in  natural one-to-one
correspondence with time-reparamatrization orbits of the space
of Hill operators $\partial_t^2+\omega^2(t)$, a problem studied
and fully solved by Kirillov \cite{Kirillov,Guieu} in terms of 
coadjoint orbits of the centrally extended Virasoro algebra.
The explanation behind this
remarkable fact is that the Hill equation $\ddot{x}(t)+\omega^2(t)
x(t)=0$ is the semi-classical problem associated to $\cal S$. It
turns out that all properties studied by Kirillov have an exact counterpart in the study of Schr\"odinger
operators, with Kirillov's orbital data in one-to-one correspondence with
Pinney-Milne dynamical invariants studied in a more physical
context. This has brought about a full understanding of the dynamical properties of time-dependent harmonic oscillators.

Consider first the Hill problem $(\partial_t^2 +\omega^2)Y(t)=0$ associated in the semi-classical limit to the Schr\"odinger operator  $\mathcal{S}$. Its monodromy matrix  $\mathbb{M}(\omega)\in SL(2,\R)$ is defined through
\be\label{mon-matrix}
\mathbb{M}(\omega)\, Y(t)= Y(t+2\pi); \qquad Y(t)\equiv\bV y_1(t)\\ y_2(t)\eV
\ee
where $y_{1,2}(t)$ are a pair of linearly independent solutions of the associated Hill's equation $\DEE{y}_{1,2}(t)+\omega^2(t)\, y_{1,2}(t)=0$. A different choice of bases leads to a conjugated monodromy matrix having the same trace. In particular, from Floquet's theory together with the orbit theory of $SL(2,\R)$, we conclude that the Hill operator $(\de_t^2+\omega^2)$ is stable (in the sense that all its solutions are bounded) if $|\Tr(\mathbb{M}(\omega))|<2$. We refer to this case as to the \IT{elliptic monodromy} case since the monodromy matrix is conjugated to a rotation. Otherwise, if $|\Tr(\mathbb{M}(\omega))|>2$ the solutions of the Hill's equation are unbounded and the matrix is conjugated to a Lorentz shift: this is  \IT{hyperbolic monodromy} case. The boundary situation $|\Tr(\mathbb{M}(\omega))|=2$ is called  \IT{unipotent monodromy} since the matrix is conjugated to a unipotent one and in general the Hill equation admits both bounded and unbounded solutions.

From a second perspective \cite{Kirillov}, the classification of the Hill operators can be done considering the so-called stabilizer group:
\be
\text{Stab}(\omega)\equiv\Big\{ \varphi(t)\in \text{Diff}_{+}(\R/2\pi\Z) \,\, : \,\, \varphi^* (\de_t^2+\omega^2)= \de_t^2+\omega^2\Big\}
\ee
consisting of all the time reparametrizations $\varphi(t)$ whose action leave the Hill operator invariant. The function $\xi(t)\in C^{\infty}(\R/2\pi\Z)$ belongs to the Lie algebra of $\text{Stab}(\omega)$ if and only if it satisfies the equation:
\be\label{KIR}
\frac{1}{2}\DEEE{\xi}(t)+2\omega^2(t)\,\DE{\xi}(t)+2\omega(t)\, \DE{\omega}(t)\, \xi(t)=0\,\, \Leftrightarrow\,\, \xi \in \text{Lie}(\text{Stab}(\omega))
\ee
and $I=\DEE{\xi}\xi-\frac{1}{2}(\DE{\xi})^2+2\omega^2\xi^2$ is a constant of motion. Generically, $\text{Lie}(\text{Stab}(\omega))$ is one-dimensional, so $\xi$ is fixed up to a multiplicative constant, and the sign of $I$ is therefore unambiguous. As follows from Kirillov's work, the type of monodromy of the system is given by the sign of this constant (assuming $\xi$ to be a real function): $I>0$ elliptic, $I=0$ unipotent,  $I<0$ hyperbolic respectively.

As shown e.g. in \cite{KW}, see also \cite{JER},
\S 2.2, the two perspectives  are actually essentially equivalent, since orbits are characterized
by the conjugacy class of the monodromy matrix and an integer called winding number.
Quite remarkably,  the equation $\eqref{KIR}$ is nothing but the derivative of the Pinney equation $\eqref{Pinney}$ in terms of the variable $\xi\equiv\zeta^2$, which allows in \cite{JER} to 'quantize' the monodromy matrix into a monodromy operator characterizing the stroboscopic evolution of wave functions of the associated Schr\"odinger operator; here we shall  be content with using the Pinney-Milne theory in connection with Kirillov's theory to consider the stroboscopic time-evolution of more directly
accessible physical quantities, like the density, etc.
The classification of the orbits \cite{Kirillov}, excluding the unipotent case, can be summarized as follows:
\begin{itemize}
\item \textit{Case 1:}  The function $\xi$ is conjugated by a time reparametrization $t \mapsto \varphi(t)$, $\varphi\in \text{Diff}_+(\R/2\pi\Z)$ to a (non-zero) constant $a\de_t$ ($a\neq 0$) which stabilizes the operator $\de^2+\alpha$, for a certain constant $\alpha>0$. The invariants are positive $I=2\alpha a^2$ and $\frac{1}{2\pi} \int_0^{2\pi} \frac{dt}{\xi(t)}=1/a$. From these defining relations one easily obtains the parameter of the orbit, $\alpha$, from the knowledge of $\xi$.  \\
\\
\item \textit{Case 2:} The function $\xi$ is conjugated to $a\sin(nt)(1+\alpha\sin(nt))\de_t$, $n=1,2,\dots$, $0\leq \alpha <1$, which stabilizes the operator $\de^2 +v_{n,\alpha}$, where
\be
v_{n,\alpha}(t)\equiv \frac{n^2}{4} \Big(\frac{1+6\alpha\sin(nt)+4\alpha^2\sin^2(nt)}{(1+\alpha\sin(nt))^2}\Big).
\ee
The monodromy matrix is hyperbolic and the invariants take the values $I=-2a^2n^2<0$. The integral of $1/\xi$ over a period  reads p.v.$ \int_{\gamma} \frac{dt}{\xi(t)}=\frac{2\pi\alpha }{a\sqrt{1-\alpha^2}}$  \footnote{see \cite{JER}, Eq. $(2.17)$}.  Knowing $\xi$ (hence $I$), it is easy to determine the parameters $n,\, \alpha$ of the orbit of $\omega$. Namely, the number of zeros of the function $\xi$ on a period is equal to  $2n$, while the defining relations for $I$ and the integral of $\xi$ over a period yield $a$ and then $\alpha$.  
\end{itemize}
For definitess,  we fix $I=2\omega_0^2>0$. This leads us in the hyperbolic case to choose instead $\xi$ to be a purely imaginary function. 

The paper is organized as follows: in the next section we present the model (Tonks Girardeau with periodic harmonic trap), its mapping to a Fermi system and its instantaneous diagonalization reducing the problem in the thermodynamic limit to a time-dependent harmonic oscillator. In section \ref{Classification} we explain the methods used in the article: we construct the time-evolved one-particle wave function through the Ermakov-Lewis dynamical invariants and we present the classification of harmonic Hamiltonians in which the frequency $\omega(t)$ is varied as a square wave. Tuning the parameters of the square-wave frequency, we show that the system explores regions with different monodromies, elliptic and hyperbolic.
Physical results for the stroboscopic evolution of the energy and of the particle density are derived in section \ref{ellipticsec} and section \ref{hyperbolicsec}, for the case of elliptic and hyperbolic monodromy respectively.  Finally, a brief summary is given in the last section.

\section{Setup}
\subsection{The model}\label{model}
In this study we describe the time evolution of a set of bosons on a lattice in the presence of an external time-dependent potential $V(t)$. The dynamics of such a model is generated by the one dimensional Bose-Hubbard model \cite{BH1,BH2}, given by
\be
\mathcal{H}(t)= -\frac{J}{2} \sum_{j=0}^L [a^\dagger_{j+1} a_j + h.c.] + \frac{U}{2}\sum_{j=0}^L n_j (n_j-1) + \sum_{j=0}^L V_j(t) n_j
\ee 
where $a^\dagger,a$ are standard bosonic operators and $n_j=a^\dagger_j a_j$ is the bosonic occupation number at site $j$. 
The kinetic coupling $J$ is set to one in the following. The on-site interaction is repulsive and modeled with a positive coupling constant $U>0$. The last term of the Hamiltonian describes the interaction of the system with the external time-dependent harmonic trap \cite{Scopa}
\be
V_j(t)= \frac{1}{2}\omega^2(t)\, j^2 -\mu
\ee
where the shift $\mu$ can be interpreted as a chemical potential and where the frequency $\omega(t)>0$ is a $2\pi$-periodic function. 
In the following, we consider only the limit of hard-core bosons  $U\gg1$ \cite{TG1,TG2,TG3} that provides a good effective description of low-density gases, $-1<\mu< 1$. In this case, the dynamics can be described in terms of a new Hamiltonian:
\be
\mathcal{H}(t)= -\frac{1}{2} \sum_{j=0}^L [b^\dagger_{j+1} b_j + h.c.] + \sum_{j=0}^L V_j(t) n_j
\label{TGH}
\ee  
with a set of operators $b^\dagger, b$ commuting for different sites and satisfying the on-site anti-commutation relations $\{b^\dagger_j,b_j\}=1$, $\{b_j,b_j\} = \{b^\dagger_j,b^\dagger_j\}=0 $ implying that the on-site occupation operator $n_j=b_j^\dagger b_j$ has only zero and one eigenvalues (hard core constraint), that is preventing multiple occupancy on a given site. The Hamiltonian $\eqref{TGH}$ can be easily mapped to a spinless tight-binding Fermi system through a Jordan-Wigner transformation \cite{JordanWigner}. Introducing the lattice fermionic operators $c_j^{\dagger}=\prod_{i<j}(1-2n_i)b^{\dagger}_j$, $c_j=(c^{\dagger}_j)^{\dagger}$, for a finite size lattice with open boundary conditions one has
\be\label{A}
\mathcal{H}(t)=\sum_{i,j=-L/2}^{L/2} c_i^{\dagger} A_{i,j}(t) c_j \; ,
\ee
 where
 \be
 A_{i,j}(t)\equiv V_i(t)\delta_{i,j} -\frac{1}{2}(\delta_{i,j+1}+\delta_{i+1,j})\; .
\ee
Notice that the hard core bosonic occupation number operator $n_j=b^{\dagger}_jb_j$ is also given by the fermionic one $c^{\dagger}_jc_j$. 

At a fixed time $t$ the quadratic Hamiltonian $(\ref{A})$ is reduced to a free theory:
\be
\mathcal{H}(t)= \sum_{q=0}^L \E_q(t) \eta^{\dagger}_q(t)\eta_q(t)\; ,
\label{Ham}
\ee
with single particle energies  $\E_q(t)$ and  where the diagonal Fermi operators $\eta^\dagger$, $\eta$ are 
related to the lattice Fermi operators $c^\dagger$, $c$  through the unitary transformation 
\be
\eta^{\dagger}_q(t)=-\sum_{i=-L/2}^{L/2} \phi_q(i,t) c_i^{\dagger}\; , \quad \eta_q(t)=-\sum_{i=-L/2}^{L/2} \phi^*_q(i,t) c_i
\label{eta}
\ee
with Bogoliubov coefficients $\phi_q(i,t)$ satisfying  $\sum_i \phi^*_q(i,t)\,\phi_p(i,t)=\delta_{qp}$.

\subsection{Instantaneous diagonalization}
\begin{figure}
\centering
\includegraphics[scale=0.4]{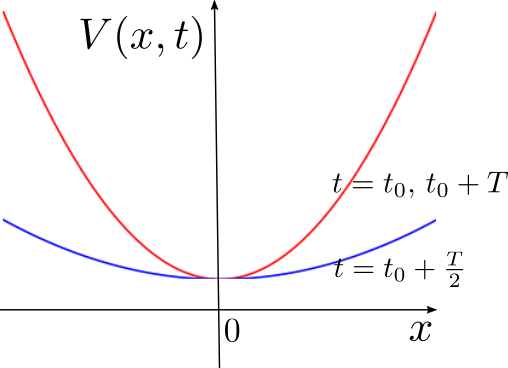}
\caption{Sketch of the driving protocol for the confining potential $V(x,t)=1+ \frac{1}{2}\omega^2(t) x^2$ as a function of the space coordinate $x$ at the initial time $t_0$, after half the period $t_0+ T/2$ and after a full period $t_0+T$. 
}\label{protocol}

\end{figure}

In the thermodynamic limit, where the lattice site $j$ is replaced by a continuous variable $x$, the instantaneous eigenvalue problem 
\be\label{eigendiscrete}
A(t)\phi_q(t)=\E_q(t) \phi_q(t)\; ,
\ee
with the choice $\mu=-1$\footnote{We are interested in the low-density regime which is set in the neighborhood of $\mu=-1$, see \cite{VIC}, since this value corresponds in the absence of the trapping potential to the transition point between the trivial Mott phase with zero density for $\mu<-1$ and the superfluid phase for $|\mu|<1$.}, reduces to a harmonic oscillator \cite{Scopa}
\be
 \frac{1}{2}(-\de_x^2+\omega^2(t)x^2)\phi_q(x,t)=\E_q(t)\phi_q(x,t)\; 
\ee 
with explicit solutions
\be
\label{phi}
\phi_{q}(x,t)= \sqrt{\frac{\sqrt{\omega(t)}}{2^q q! \sqrt{\pi}}} e^{-\frac{\omega(t)}{2}x^2} \; {\rm He}_q(x\sqrt{\omega(t)})\;, \qquad 
\E_{q}(t)= \omega(t)(q+\frac{1}{2})\;, 
\ee
where $q\in \mathbb{N}$ and ${\rm He}_q$ denotes the $q^{th}$ Hermite polynomial with physical normalization. In Fig.\ref{protocol}, an illustration of the confining trap in the thermodynamic limit is shown.
Notice that the  lattice structure has been removed in the thermodynamic limit and the model reduces to a Lieb-Liniger model \cite{LL} with high-repulsive interactions. 
\subsection{Initial conditions}
The system is composed of $N$ bosons initially (at $t=0$) assumed to be in the ground state of $\mathcal{H}(0)$ which is simply given
thanks to \eqref{Ham} and \eqref{phi} by
\be
\label{GS0}
|\GS(0)\rangle = \prod_{q=0}^{N-1} \eta^{\dagger}_q(0)| 0\rangle \; , 
\ee
where $|0\rangle$ is the vacuum state such that $\eta_q(0)| 0\rangle = 0$ $\forall q$. The associated energy  of the initial state is thus
\be
E_{\GS}(0) = \sum_{q=0}^{N-1} \E_q(0)\; .
\ee
\section{Dynamical invariants and classification of the harmonic Hamiltonians}\label{Classification}
\subsection{Time-evolution of the one-particle wave function}
Let us consider the  single particle Schr\"odinger equation 
\be
i\de_t \,\psi_k(x,t)= \frac{1}{2}\left(-\de_x^2+\omega^2(t)x^2\right) \psi_k(x,t)
\ee 
governing the time evolution of the initial $k^{th}$ eigenstate  $\psi_k(x,0)= \phi_k(x)\equiv \phi_k(x,0)$. 
 A convenient way to get a solution of the one-particle Schr\"odinger equation is to expand the wave function in terms of the eigenvectors of the Ermakov-Lewis (EL) operator $\EL$ which is given for harmonic Hamiltonians by \cite{EL1}:
\be\label{EL}
\EL(x,t)\equiv \frac{1}{2}\Big(\frac{\omega_0^2\,x^2}{\zeta^2(t)}-(\zeta(t)\, \de_x-i\dot{\zeta}(t)\,x )^2\Big)
\ee
where $\omega_0\equiv\omega(0)$ and where $\zeta$ is a solution of the Pinney equation \cite{PINNEY}
\be\label{Pinney}
\DEE{\zeta}(t)+\omega^2(t)\, \zeta(t)=\omega_0^2\, \zeta^{-3}(t)\, .
\ee
One can easily prove that this operator is a dynamical invariant, $\frac{d}{dt}\EL=\de_t \, \EL -i[\EL , \Ha]=0$, see e.g. \cite{EL2}. The expansion  of the one-particle wave function on the basis of the eigenfunctions $h_{\lambda}$ of the EL operator is given by \cite{EL2}
\be\label{onewave}
\psi_k(x,t)= \sum_{\lambda \in \, \text{spec}(\EL)} c_{k,\lambda}\,\, e^{i\alpha_{\lambda}(t)} \, h_{\lambda}(x,t),
\ee
where $c_{k,\lambda}\equiv\braket{h_{\lambda}(0)|\phi_k}$ are the overlap coefficients and where the dynamical phases $\alpha_\lambda$ are given by the solution of the equation
\be\label{alpha}
\frac{d}{dt}\alpha_{\lambda}(t)=\braket{\DE{h}_{\lambda}(t)|(i\de_t-\Ha)h_{\lambda}(t)}\; 
\ee
with $\alpha_\lambda(0)=0$. 
One can notice that the time-evolution of the one-particle wave function in the EL bases $\eqref{onewave}$ is merely  a gauge transformation of the EL eigenvectors.

\subsection{Classification of the harmonic Hamiltonians}
We denote by $\mathcal{S}$ the Schr\"odinger operator $\mathcal{S}\equiv i\de_t-\Ha$ associated to the single particle Hamiltonian $\Ha=1/2(-\partial_x^2 +\omega^2(t)x^2)$  with a periodic time-depedent frequency $\omega(t)=\omega(t+2\pi)$.
As explained in the introduction, 
Schr\"odinger operators in the same orbit are connected by a time reparametrization $t\To \varphi(t)$. This implies that if we know the one-particle wave function $\psi_{1}(x,t)$  for a certain function $\omega_{1}(t)$ then the one-particle wave function $\psi_{2}(x,t)$ for an $\omega_{2}(t)$ belonging to the same orbit as $\omega_{1}(t)$ is simply given by \cite{JER}
\be
\psi_{2}(x,t)=(\DE{\varphi})^{-1/4} \, \exp(\frac{i}{4}\frac{\DEE{\varphi}}{\DE{\varphi}}\, x^2)\, \psi_{1}(x\, (\DE{\varphi})^{-1/2},\varphi)
\ee
while Ermakov-Lewis invariants $\xi_1=\zeta_1^2$ and  $\xi_2=\zeta_2^2$ are related through
\be
\zeta^2_{2}(t)=(\DE{\varphi})^{-1}\, \zeta^2_{1}(\varphi^{-1})\; .
\ee

Elliptic and hyperbolic  monodromy classes can be generated by the simple situation where 
the system is subjected to a periodic time-dependent trap whose frequency is varied as a square wave 
\be\label{piece}
\omega(t)=\omega(t+2\pi)=\begin{cases} \omega_1 \qquad \text{if}\qquad 0 < t < \tau\\  \omega_2 \qquad \text{if}\qquad \tau < t < 2\pi
\end{cases}\ee
where $\omega_1,\omega_2>0$ are constants. For this setting, the Eq. $\eqref{Pinney}$ becomes in terms of the variable $\xi\equiv \zeta^2$
\be
\DEE{\xi}(t)\, \xi(t)-\frac{1}{2} (\DE{\xi}(t))^2+2\omega^2(t)\, \xi^2(t)=2\omega_1^2\; .
\ee
Away from the discontinuities ($t\neq0^{\pm}, \, \tau^{\pm}$), we can differentiate the last equation obtaining
\be
\DEEE{\xi}(t)+4\omega^2_{1\Bs2} \, \DE{\xi}(t)=0\; ,
\ee
which can be readily integrated 
\be\label{xi-piece}
\xi(t)=\begin{cases} \xi_1(t)=\alpha_1\, e^{i2\omega_1 t}+\gamma_1+\beta_1\, e^{-2i\omega_1t}\;, \qquad 0<t<\tau\\
\xi_2(t)=\alpha_2\, e^{i2\omega_2 t}+\gamma_2+\beta_2\, e^{-2i\omega_2t}\; , \qquad \tau<t<2\pi\\
\end{cases}\ee
with $\alpha_{1\Bs2},\, \gamma_{1\Bs2},\, \beta_{1\setminus2}$ constants that have to be fixed imposing continuity conditions at $t=0$ and $t=\tau$. To do so, we can relate the unknown constants to the function $\xi$ and its derivatives using the notation:
\be\label{xi-coef}
\vec{\xi}(t)\equiv \bV \xi(t) \\ \DE{\xi}(t)\\ \DEE{\xi}(t) \eV = \mathbb{A}(t) \, \bV \alpha_{1\Bs2} \\ \gamma_{1\Bs2} \\ \beta_{1\Bs2} \eV
\ee
where
\be
\mathbb{A}(t)\equiv \bV e^{i2\omega(t)t} & 1 & e^{-i2\omega(t)t} \\ i2\omega(t)\,  e^{i2\omega(t) t} &0&  -i2\omega(t)\,e^{-i2\omega(t) t}\\ -4\omega^2(t)\, e^{i2\omega(t) t}&0& -4\omega^2(t) \,  e^{-i2\omega(t) t} \eV\; , \qquad t\in (0,\tau)\vee(\tau,2\pi)\; .
\ee
Imposing the continuity of the function $\xi$ and of its first derivative at $t=0,\, \tau$ we have the conditions:
\begin{subequations}\label{cont}
\be
\vec{\xi}(\tau^+)=B_{[\tau^+\leftarrow \tau^-]} \, \vec{\xi}(\tau^-),\qquad B_{[\tau^+\leftarrow \tau^-]}\equiv \bV 1&0&0\\ 0&1&0\\ -2(\omega_2^2-\omega_1^2)&0&1\eV\; ,
\ee
and
\be
\vec{\xi}(0^+)=B_{[0^+\leftarrow 2\pi^-]} \, \vec{\xi}(2\pi^-),\qquad B_{[0^+\leftarrow 2\pi^-]}\equiv \bV 1&0&0\\ 0&1&0\\ -2(\omega_1^2-\omega_2^2)&0&1\eV\; .
\ee
\end{subequations}
The set of conditions $\eqref{cont}$ together with $\eqref{xi-coef}$ yields 
\begin{subequations}\label{eigenproblem}
\be
\bV \alpha_2 \\ \gamma_2 \\ \beta_2 \eV = \underbrace{ \mathbb{A}^{-1}(\tau^+) \, B_{[\tau^+\leftarrow \tau^-]}\, \mathbb{A}(\tau^-)}_{\equiv\mathbb{T}_1} \, \bV \alpha_1 \\ \gamma_1 \\ \beta_1 \eV\; ,
\ee
and
\be
\bV \alpha_1 \\ \gamma_1 \\ \beta_1 \eV = \underbrace{\mathbb{A}^{-1}(0^+) \, B_{[0^+\leftarrow 2\pi^-]}\, \mathbb{A}(2\pi^-)}_{\equiv\mathbb{T}_2} \, \bV \alpha_2 \\ \gamma_2 \\ \beta_2 \eV\; .
\ee
\end{subequations}
Thus 
\be
\bV \alpha_1 \\ \gamma_1 \\ \beta_1 \eV = \mathbb{T}_2\mathbb{T}_1 \bV \alpha_1 \\ \gamma_1 \\ \beta_1 \eV 
\ee
is  the eigenvector of $\mathbb{T}_2\, \mathbb{T}_1$ associated  to the eigenvalue $1$. From the first set of coefficients in the region $t\in[0,\tau]$, we can extract then the second set using the first relation in $\eqref{eigenproblem}$.  

\begin{figure}
\centering
\includegraphics[scale=0.3]{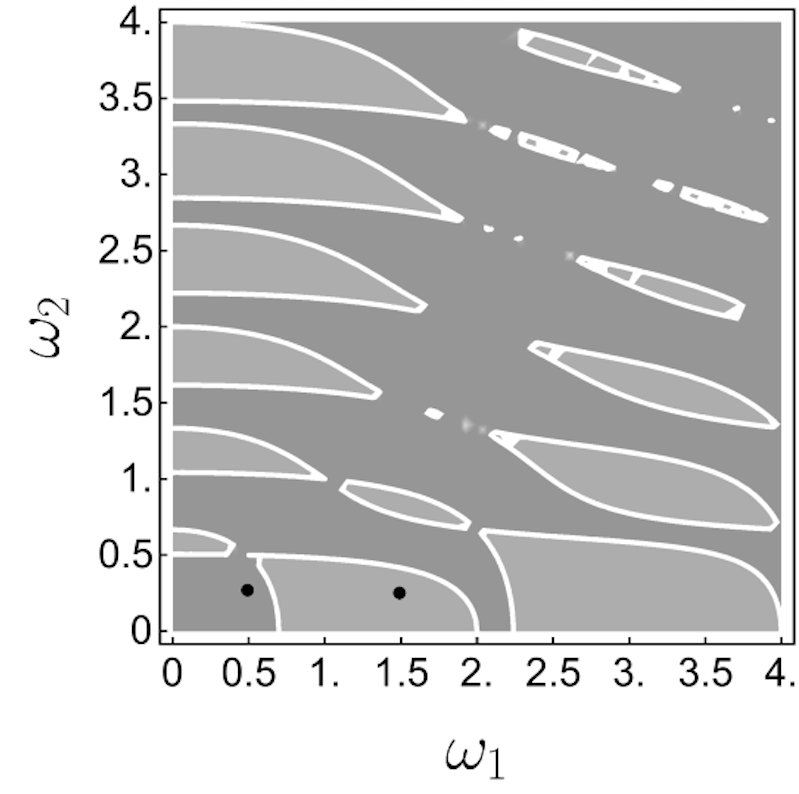}
\caption{The two different classes, elliptic in dark grey and hyperbolic in light grey,  as obtained from the half trace of the monodromy matrix in $\eqref{mon}$ for $\tau= \pi/2$ as a function of $\omega_1$ and $\omega_2$. The black dots represent the values of $(\omega_1,\omega_2)$ considered explicitly below in the computation of the physical quantities. }\label{monodromy}
\end{figure}

\begin{figure}
\centering
\includegraphics[scale=0.25]{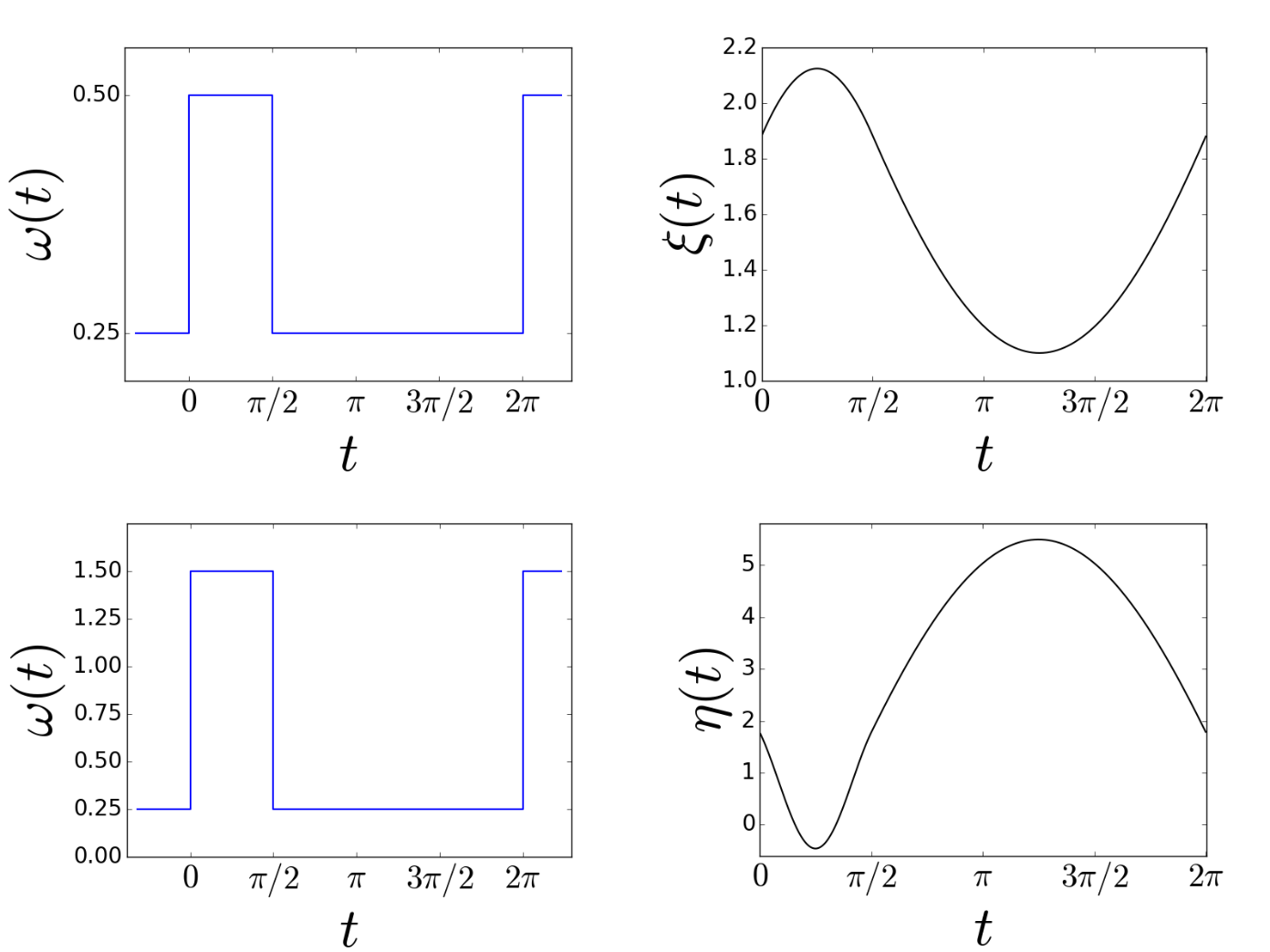}
\caption{The square-wave frequency $\eqref{piece}$ \textsl{(left)} and the associated solution \textsl{(right)} of the Eq.$\eqref{Pinney}$ for different values of the parameters: \textsl{(Top)} Elliptic case with $\omega_1=0.5,\, \omega_2=0.25,\, \tau=\pi/2$, \textsl{(Bottom)} hyperbolic  case with $\omega_1=1.5,\, \omega_2=0.25,\, \tau=\pi/2$ where the associated solution is purely imaginary ($\xi=i\eta$).}\label{PIECE}
\end{figure}

The classification  into elliptic and hyperbolic cases is obtained from the trace of the monodromy matrix $\mathbb{M}$ of the associated Hill problem (see \ref{piece-mon}):
\be\label{mon}
\frac{1}{2}\text{Tr}(\mathbb{M}(\omega_1,\omega_2,\tau)) = \cos(\omega_1\tau) \cos(\omega_2(2\pi-\tau)) -\frac{1}{2}\Big(\frac{\omega_1}{\omega_2}+\frac{\omega_2}{\omega_1}\Big) \sin(\omega_1\tau) \sin(\omega_2(2\pi-\tau))\; .
\ee
The elliptic situation arises if $|\text{Tr}(\mathbb{M} )| <2$ and the hyperbolic case for $|\text{Tr}(\mathbb{M}) | >2$.
In Fig. \ref{monodromy} we show the elliptic and hyperbolic domains for a given value of $\tau$. Note in particular that if $\omega_1=\omega_2\equiv\omega$,  $|\Tr(\mathbb{M})|= |2\cos\omega|\leq 2 $, as expected.
In Fig.\ref{PIECE} we show the solutions $\xi$ for two different points of the monodromy phase diagram, $(\omega_1,\omega_2)=(1/2,1/4)$ and $(\omega_1,\omega_2)=(3/2,1/4)$, corresponding respectively to the elliptic and hyperbolic case. 

%
%
%
%
%
%
%
%
%
%
%
%
%
%
%
%
%
%
\section{The case of elliptic monodromy}\label{ellipticsec}
\subsection{Discussion}
Consider a point in the plan $(\omega_1,\omega_2)$ for which  the system belongs to an orbit of elliptic monodromy. In this case the EL operator $\eqref{EL}$ takes the form of an harmonic oscillator. Its eigenfunctions are given by \cite{RogUnt}
\be
h_{\lambda}(x,t)=\frac{1}{\sqrt{\zeta(t)}} \, \exp(i\frac{\DE{\zeta}(t)}{2\zeta(t)}\, x^2) \, \phi_{\lambda}(\frac{x}{\zeta(t)})
\ee
and the spectrum is $\omega_0(\lambda+1/2)$ with  $\lambda=0,1,2,...$ natural integer. The dynamical phase $\eqref{alpha}$ acquired by the eigenvectors is \cite{RogUnt}
\be
\alpha_{\lambda}(t)=-\omega_0(\lambda+\frac{1}{2})\int_0^t \frac{dt'}{\zeta^2(t')}\; .
\ee
All the information about the choice of the periodic frequency function $\omega(t)$ is thus contained in the $2\pi$-periodic function $\zeta(t)$, which is a periodic solution of the Pinney equation $\eqref{Pinney}$ (it has been proved in \cite{periodiczeta} that the Pinney equation $\eqref{Pinney}$ always admits a periodic solution if $\omega$ is periodic).

Notice that the constant frequency $\omega(t)=\omega_0$ situation, for which a solution of $\eqref{Pinney}$ is given by the constant $\zeta\equiv 1$, also belongs to the elliptic case. We refer to this case as the equilibrium (or adiabatic) limit since it is conceptually equivalent to a system which adapts itself to the instantaneous value of the trap during the time evolution. 
\subsection{Stroboscopic evolution of the one-particle wave function}
The stroboscopic dynamics of the one-particle wave function is obtained by considering a time-step increase of one period $\Delta t=2\pi$. From $\eqref{onewave}$ we obtain after $n\in \mathbb{N}$ periods:
\be\label{Eonewave1}
\psi_k(x,2\pi n)= \sum_{\lambda=0}^{\infty} \int_{\R} dy\,\, h^*_{\lambda}(y,0) \, \phi_k(y) \,\,  e^{-inT(\lambda+\frac{1}{2})} \,\, e^{i\frac{\DE{\zeta}_0}{2\zeta_0}\, x^2} \frac{1}{\sqrt{\zeta_0}}\, \phi_{\lambda}(\frac{x}{\zeta_0})\; ,
\ee
where we have introduced the notations $\zeta_0\equiv\zeta(0)$, $\DE{\zeta}_0\equiv \DE{\zeta}(0)$ and we have defined
\be
T\equiv\omega_0\int_0^{2\pi} \frac{dt'}{\zeta^2(t')} \; .
\ee
Using the definition of the Mehler kernel $\K$ (see \ref{Mehler}), the expression $\eqref{Eonewave1}$ can be written in the form
\be\label{EonewaveF}
\psi_k(x,2\pi n)= \int_{\R} dy\, \exp(i\frac{\DE{\zeta}_0}{2\zeta_0}\,(x^2-y^2))\,\, \phi_k(y) \,\, \K(\overline{x},\overline{y}|inT)\; ,
\ee
where the kernel is explicitly given by
\be\label{Kell}
\K(\overline{x},\overline{y}|inT)= \frac{\sqrt{\omega_0}}{\zeta_0} \frac{\exp(\frac{i}{2}(\overline{x}^2+\overline{y}^2)\cot(nT)-i\overline{x}\,\overline{y}/\sin(nT))}{\sqrt{2\pi i \, \sin(nT)}}
\ee
and we have introduced the useful notation $\overline{x}\equiv x\sqrt{\omega_0}/\zeta_0$, $\overline{y}\equiv y\sqrt{\omega_0}/\zeta_0$.
\subsection{Stroboscopic evolution of the energy spectrum}\label{Eenergy}
\begin{figure}
\centering
\includegraphics[scale=0.4]{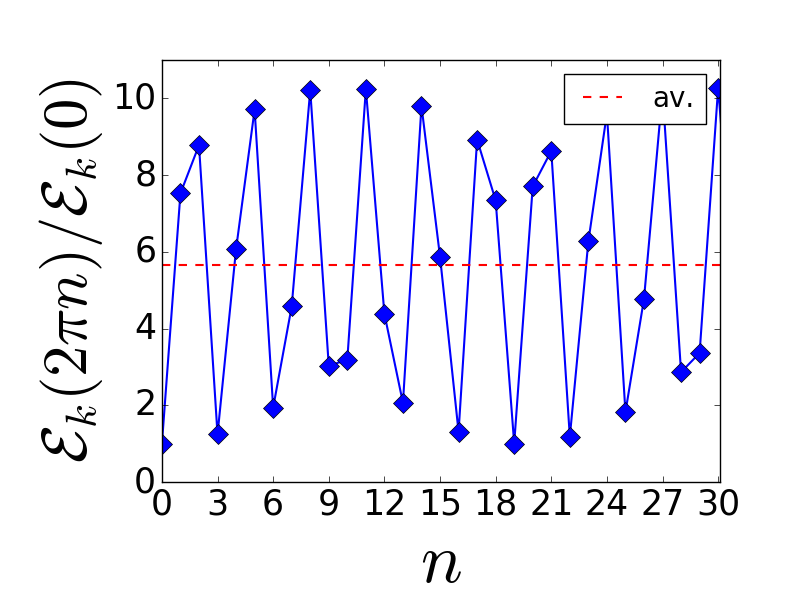}
\caption{The stroboscopic evolution of the energy $\E_k(2\pi n)/\E_k(0)$ for the elliptic point $(\omega_1,\omega_2)=(1/2,1/4)$  as function of $n$ and its average value (dashed line).}\label{EE}
\end{figure}

We want to understand how the energy spectrum of the Hamiltonian $\eqref{Ham}$ at $t=0$ evolves in time when the system is subjected to a periodic variation of the harmonic trap.
In particular, exploiting the $2\pi$-periodicity of the Hamiltonian $\eqref{Ham}$ and the result $\eqref{EonewaveF}$, we look at the energy of the $k^{th}$ level of the initial Hamiltonian after $n$ periods:
\be\label{energy}\begin{split}
\E_{k}(2\pi n)=\braket{\psi_k(2\pi n)|\Ha(0)|\psi(2\pi n)}=\\
\braket{\psi(2\pi n)|-\frac{1}{2}\de_x^2\, |\psi(2\pi n)}+\braket{\psi(2\pi n)|\frac{\omega_0^2}{2}\, x^2|\psi(2\pi n)}= \E_{k|1}(2\pi n) +\E_{k|2}(2\pi n)\; ,
\end{split}\ee
where we have divided the expectation value into the kinetic part ($\E_{k|1}$) and the potential ($\E_{k|2}$). The expectation value of the potential is given by the expression
\be\label{E2}
\E_{k|2}(2\pi n)=\int_{\R} dy\, \int_{\R} dy' \,\, e^{i\DE{\zeta}_0\,(y^2-y'^2)/2\zeta_0}\,\, \phi_k(y)\, \phi_k(y') \,\, \mathcal{I}_2(y,y'|nT)
\ee
where 
\be
\mathcal{I}_2(y,y'|nT)\equiv\frac{\omega_0^2}{2} \int_{\R} dx\,\, \K^*(\overline{x},\overline{y}|inT)\,\, x^2\,\, \K(\overline{x},\overline{y}'|inT)\; .
\ee
Performing the integration over $x$ of the kernel $\mathcal{I}_2$ and inserting the expression in $\eqref{E2}$ we obtain
\be\bspl
\E_{k|2}(2\pi n)=\frac{1}{2}\zeta_0^4 \sin^2(nT)\, \int_{\R} dy\, \int_{\R} dy'\, \delta(y-y')\\
\{-\frac{1}{2}(\de_y^2+\de_{y'}^2)\}\Big\{\phi_k(y)\,\, \phi_k(y')\,\, \exp(\frac{i\omega_0}{2\zeta_0^2}(-\cot(nT)+\Zz)(y^2-y'^2))\Big\}\; .
\end{split}\ee
The action of the derivative and the integral can be easily computed using the recursion properties of the Hilbert-Hermite functions $\phi_q$ \cite{Abramovitz}. The result is
\be\label{E2F}\bspl
\E_{k|2}(2\pi n)=\frac{1}{2}\Big[\zeta_0^4\sin^2(nT)+ \Big(\cos(nT)-\Zz\sin(nT) \Big)^2 \Big] \, \omega_0(k+\frac{1}{2})\; .
\end{split}\ee
With the same procedure we also compute the expectation value of the kinetic term:
\be\label{E1}
\E_{k|1}(2\pi n)=\int_{\R} dy\, \int_{\R} dy' \,\, e^{i\DE{\zeta}_0\,(y^2-y'^2)/2\zeta_0}\,\, \phi_k(y)\, \phi_k(y') \,\, \mathcal{I}_1(y,y'|nT)
\ee
where $\mathcal{I}_1(y,y'|nT)$ is defined to be  
\be\label{I1}
\mathcal{I}_1(y,y'|nT)\equiv\frac{1}{2}\int_{\R} dx\,\,  e^{-i\frac{\dot{\zeta_0}}{2\zeta_0}\; x^2} \, \K^*(\overline{x},\overline{y}|inT)\,\,(- \de_x^2)\,\, e^{i\frac{\dot{\zeta_0}}{2\zeta_0}\; x^2}\, \K(\overline{x},\overline{y}'|inT)\; .
\ee
Performing the integration over $x$ of the kernel $\mathcal{I}_1$, the expression $\eqref{E1}$ becomes
\be\bspl
\E_{k|1}(2\pi n)=\frac{\omega_0}{2\zeta_0^2}\, \int_{\R} dy \, \int_{\R} dy'\,\, \delta(y-y') \, \Big\{ \frac{\zeta_0^2}{\omega_0}\Big(\cos(nT)+(\Zz)\sin(nT)\Big)^2 \de_y\, \de_{y'}\\
-i\Big(\cot(nT)+(\Zz)\Big)[\de_y\, y'- y\, \de_{y'}] +\frac{\omega_0}{\zeta_0^2}\,\frac{ y\,y'}{\sin^2(nT)}\Big\}\\
\Big\{ \phi_k(y)\,\, \phi_k(y')\,\, \exp(\frac{i\omega_0}{2\zeta_0^2}(-\cot(nT)+\Zz)(y^2-y'^2))\Big\}\; ,
\end{split}\ee
which finally leads to
\be\label{E1F}\bspl
\E_{k|1}(2\pi n)=\frac{1}{2}\Big[ \Big(\cos(nT)+\Zz\sin(nT)\Big)^2 +\frac{1}{\zeta_0^4} \sin^2(nT) \Big(1+(\Zz)^2\Big)^2 \Big]\,\omega_0(k+\frac{1}{2}) \; .
\end{split}\ee
The stroboscopic evolution of the energy $\E_k$ is given by the sum of the kinetic part $\eqref{E1F}$ and of the potential part $\eqref{E2F}$:
\be\label{EFIN}
\E_{k}(2\pi n)= \Big[ \cos^2(nT) +(\Zz)^2 \sin^2(nT) +\frac{1}{2} \sin^2(nT) \Big(\zeta_0^4 +\frac{1}{\zeta_0^4}\Big(1+(\Zz)^2\Big)^2\Big)\Big] \; \omega_0(k+\frac{1}{2})\; .
 \ee
Notice that for $n=0$ we recover the initial value of the energy $\E_k(0)=\omega_0(k+1/2)$. Furthermore, setting $\zeta\equiv 1$ the equilibrium value of the energy is recovered at any instant of time. It is interesting to consider a time average of the energy which can be defined as follows:
\be
\braket{\E_k}\equiv\lim_{m\To\infty} \frac{1}{m}\sum_{n=0}^m \E_k(2\pi n) = \int_0^{2\pi} \frac{d\vartheta}{2\pi}\,\, \E_k(\vartheta)\; ,
\ee
where the identity above holds not only in $L^1$ as follows from standard ergodic theory arguments, see e.g. \cite{ergodicity}, but also in a stronger sense, implying in particular pointwise convergence \cite{Gerald}. The averaged energy is thus
\be\label{ellEA}
\braket{\E_k}=\Big[ 1+ (\Zz)^2 +\frac{1}{2}\Big(\zeta_0^4 +\frac{1}{\zeta_0^4} \Big(1+(\Zz)^2\Big)^2\Big)\Big] \frac{\omega_0}{2}(k+\frac{1}{2})\; .
\ee
In the special case $\dot{\zeta}_0=0$, the system exhibits a sort of equipartition between the expectation values of the kinetic part and the potential part:
\be
\braket{\E_{k|1}}=\zeta_0^4\, \braket{\E_{k|2}}
\ee
where the factor $\zeta_0^4$ can be explained by the different scaling of the kinetic part of the Hamiltonian $\eqref{Ham}$ in terms of the variable $\overline{x}=x\sqrt{\omega_0}/\zeta_0$
\be
\Ha(0)=\frac{\omega_0\zeta_0^2}{2}(-\frac{1}{\zeta_0^4}\,\de^2_{\overline{x}}+\overline{x}^2)\; .
\ee
The stroboscopic evolution of the energy for the elliptic point $(\omega_1,\omega_2)=(1/2,1/4)$ is shown in Fig.\ref{EE}. 
\subsection{Stroboscopic evolution of the $N$-particle density}\label{NpartDENS}
\begin{figure}
\centering
\includegraphics[scale=0.35]{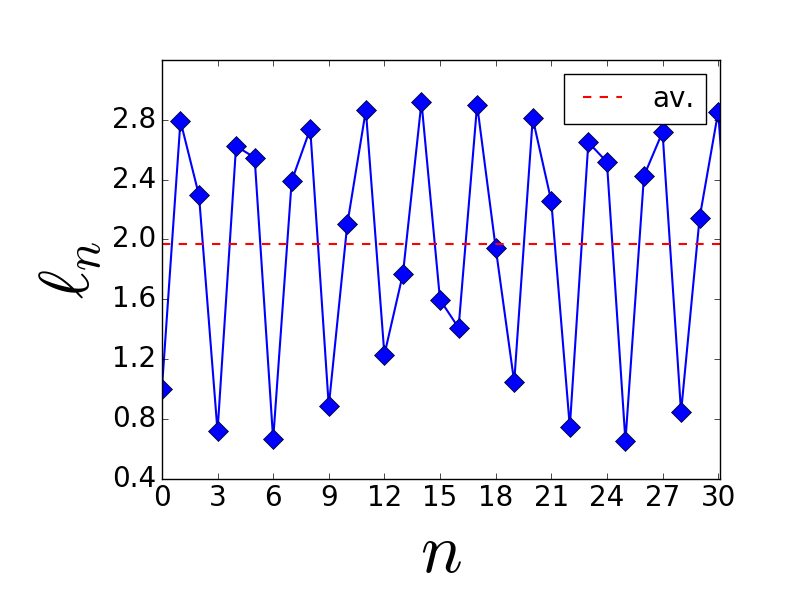} \qquad \includegraphics[scale=0.35]{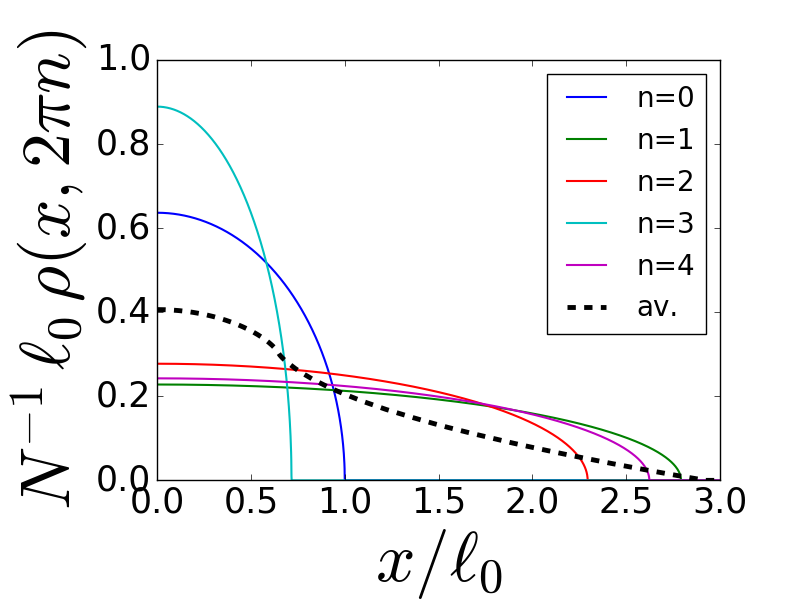}
\caption{\textsl{(Left)} The support $\ell_n$ of Eq.$\eqref{ln}$    for the elliptic point $(\omega_1,\omega_2)=(1/2,1/4)$    as function of $n\in\mathbb{N}$. \textsl{(Right)} The associated stroboscopic evolution of the rescaled $N$-particle density $\ell_0\, \rho(x,2\pi n)/N$ $\eqref{densEF}$  as function of $\widetilde{x}=x/\ell_0$. The dashed line shows the time averaged density of $\eqref{av-density}$ computed numerically with $m=10^3$.}\label{Ndens}
\end{figure}

In this section we derive the stroboscopic evolution of the particle density for a system composed of $N$ hard-core bosons initially prepared in the ground state $\eqref{GS0}$. The $N$-particle density is  given by (see \ref{Ndensity}):
\be\label{generalNpart}
\rho(x,2\pi n)=\sum_{k=0}^{N-1}|\psi_k(x,2\pi n)|^2\; .
\ee
Using the result $\eqref{EonewaveF}$ for the stroboscopic evolution of the one-particle wave function we have
\be\label{rhoN}\bspl
\rho(x,2\pi n)= \int_{\R} dy\, \int_{\R} dy' \,\, e^{i\frac{\dot{\zeta_0}}{2\zeta_0}(y^2-y'^2)}\;  \K^*(\overline{x},\overline{y}|inT) \,  \K(\overline{x},\overline{y}'|inT)\, \K_N(y,y')
\end{split}\ee
where we have introduced the Christoffel-Darboux kernel \cite{Abramovitz}
\be\label{KN}
\K_N(y,y')\equiv\sum_{k=0}^{N-1} \phi_k(y) \phi_k(y')=\sqrt{2N}\Big( \frac{\phi_N(y')\,\phi_{N-1}(y)-\phi_N(y)\,\phi_{N-1}(y')}{2\sqrt{\omega_0}(y'-y)}\Big) \; .
\ee
If we consider a large number of bosons $N\gg1$, the Hilbert-Hermite functions $\phi_N(x)$ take significant values only in the region $|x|\leq \sqrt{2N/\omega_0}$ where the zeros of the Hermite polynomials are located. Outside that region the Hilbert-Hermite functions decay exponentially fast. In this limit they can be represented by the asymptotic expansion \cite{asyHermite}
\be\label{asyN}
\phi_{N-\frac{1}{2}\pm\frac{1}{2}}(x) \overset{N\To\infty}{\simeq} \omega_0^{1/4}\, (2N)^{-1/4}\, \sqrt{\frac{2}{\pi}}\, (\sin\theta)^{-1/2}\, \sin\Big[\frac{N}{2}(\sin(2\theta)-2\theta)\mp \frac{\theta}{2}+\frac{3}{4}\pi\Big]\; ,
\ee
where $x=\sqrt{(2N/\omega_0)}\cos\theta$ and $\theta \in [0,\pi]$.
The Christoffel-Darboux kernel $\eqref{KN}$ can be handled using the asymptotic expression $\eqref{asyN}$ and becomes
\be\label{KNasy}
K_N(y,y')\overset{N\To\infty}{\simeq} \sqrt{\omega_0}\, \frac{(2N)^{-1/2}}{\pi(\cos\theta'-\cos\theta)}\, (\sin\theta\, \sin\theta')^{-1/2}\, F(\theta, \theta')\; , \qquad \begin{matrix} y=\sqrt{(2N/\omega_0)}\cos\theta \\\\ y'=\sqrt{(2N/\omega_0)}\cos\theta' \end{matrix}
\ee
where we have introduced the function
\be
F(\theta,\theta')\equiv \sin\Big[\frac{N}{2}(g(\theta')-g(\theta))\Big]\, \sin\Big(\frac{\theta+\theta'}{2}\Big)+\cos\Big[\frac{N}{2}(g(\theta')+g(\theta))\Big]\, \sin\Big(\frac{\theta'-\theta}{2}\Big)
\ee
and $g(\theta)\equiv \sin(2\theta)-2\theta$. Using this and computing explicitly the product of the two Mehler kernels in $\eqref{rhoN}$, the $N$-particle density in the limit $N\gg 1$ is obtained to be
\be\bspl
\rho(x,2\pi n)\overset{N\To\infty}{\simeq} \frac{\sqrt{2N\, \omega_0}\,\, \zeta_0^{-2}}{2\pi^2\, |\sin(nT)|} \,\int_0^{\pi}  \int_0^{\pi} d\theta \, d\theta' \, \frac{\sqrt{\sin\theta\, \sin\theta'}}{\cos\theta'-\cos\theta} \,\, F(\theta,\theta')\Big\{
\\ \exp\Big(i2N(\frac{(-\cot(nT)+\Zz)}{2\zeta^2_0}(\cos^2\theta-\cos^2\theta') +\frac{\widetilde{x}}{\zeta_0^2 \sin(nT)} (\cos\theta-\cos\theta'))\Big)\Big\}
\end{split}
\ee
where $\widetilde{x}\equiv x/ \sqrt{(2N/\omega_0)} $ and with the conditions $|y|,|y'| \leq |x|$. 
The last expression is a double oscillatory integral. Applying the stationay phase method, we conclude that the phase factors contributes with subleading terms $\sim\mathcal{O}(N^{-1/2})$ while the leading contribution $\sim \mathcal{O}(\sqrt{N})$ comes from the region in which $\theta' \simeq \theta$. By Taylor expanding the functions around  $\theta'= \theta+\delta$, $\delta\ll 1$, we obtain
\be\label{rhoEXP}\bspl
\rho(x, 2\pi n)\overset{N\To\infty}{\simeq} \frac{\sqrt{2N\,\omega_0}\,\, \zeta_0^{-2}}{2\pi^2\, |\sin(nT)|} \,\int_0^{\pi} d\theta \, \int_{\theta-\delta}^{\theta+\delta} \frac{ d\theta' }{\theta'-\theta} \ \Big\{ \\
\sin\Big[N(\cos(2\theta)-1)(\theta'-\theta)\Big]\,\sin\theta + \cos\Big[N\, g(\theta)\Big]\, \sin\Big(\frac{\theta'-\theta}{2}\Big) \Big\} \\
 \exp\Big(i2N(\frac{(-\cot(nT)+\Zz)}{2\zeta_0}\,\sin(2\theta)(\theta'-\theta) +\frac{\widetilde{x}}{\zeta_0^2\, \sin(nT)} \, \sin\theta (\theta'-\theta))\Big)\; .
\end{split}\ee
The estimation of the integral
\be
\int_{|\gamma|>\delta} d\gamma\,\,  \frac{\sin(N\, \gamma)}{\gamma}\sim \mathcal{O}(N^{-1}\delta^{-1})\To 0\,, \qquad \delta\gg N^{-1}
\ee
allow us to integrate the expression $\eqref{rhoEXP}$ over $\theta'$ using the well-known integral $\int_{\R} dx\, \sin(qx)/x=\pi\, \sgn(q)$, $q\in\R$, obtaining
\be 
\rho(x,2\pi n) \overset{N\To\infty}{\simeq}  \frac{\sqrt{2N\,\omega_0}\,\, \zeta_0^{-2}}{4\pi\, |\sin(nT)|} \,\int_0^{\pi} d\theta \, \sin\theta \,\, (\sgn(\Phi_+)+\sgn(\Phi_-))\; ,
\ee
where 
\be
\Phi_{\pm}(\theta)\equiv  \sqrt{1+a^2} \sin\Big[\theta \mp \arcsin(\frac{a}{\sqrt{1+a^2})}\Big]\mp b\,\, , \qquad \begin{matrix} a\equiv(-\cot(nT)+\Zz)/\zeta_0^2 \\ \\ b\equiv\widetilde{x}/(\zeta_0^2\sin(nT))\; . \end{matrix}
\ee
Notice that the functions $\Phi_{\pm}$ have roots only if the condition
\be\label{ln}
|\widetilde{x}|\leq \ell_n\equiv\sqrt{\zeta_0^4\,\sin^2(nT) +\Big(\Zz\sin\, (nT)-\cos(nT)\Big)^2}
\ee
is satisfied. Otherwise the function $\Phi_{\pm}$ have opposite signs and there are no contribution for the $N$-particle density. Inside the support $|\widetilde{x}|\leq \ell_n$ the study of the sign of the functions $\Phi_{\pm}$ leads to the result:
\be\label{densEF}
\rho(x,2\pi n) \overset{N\To\infty}{\simeq} \frac{2}{\pi}\, \frac{N}{\ell_0\, \ell_n} \, \sqrt{1-\frac{\widetilde{x}^2}{\ell^2_n}}
\ee
with $\ell_0\equiv\sqrt{2N/\omega_0}$ the typical length scale of the system at the equilibrium \cite{Scopa}. Indeed, setting $\zeta\equiv 1$ the well-known semi-circle law is recovered. The stroboscopic evolution of the $N$-particle density  for the elliptic point $(\omega_1,\omega_2)=(1/2,1/4)$  and the associated dynamical support $\ell_n$ are shown in Fig.\ref{Ndens}.
The time-averaged density is given by
\be\label{av-density}
\braket{\rho(x)}\overset{N\To\infty}{\simeq}\lim_{m\To\infty} \frac{2}{\pi}\, \frac{N}{\ell_0\, m} \sum_{n=0}^m  \frac{1}{\ell_n} \, \sqrt{1-\frac{\widetilde{x}^2}{\ell^2_n}}
\ee
and can be easily computed numerically. As one can see in  Fig.\ref{Ndens} it deviates significantly from a simple semi-circle law. 
%
%
%
%
%
%
%
%
%
%
%
%
%
%
%
%
%
%
\section{The case of hyperbolic monodromy}\label{hyperbolicsec}
\subsection{Discussion}
The EL operator $\eqref{EL}$ in terms of the variable $\xi\equiv\zeta^2$ takes the form
\be
\EL(x,t)=\frac{1}{2\xi(t)}(\omega_0^2\,x^2+(i\xi(t)\,\de_x+\frac{1}{2}\DE{\xi}(t)\,x)^2) \; .
\ee
If we set the frequency $\omega(t)$ in such a way to have hyperbolic monodromy, the function $\xi\in i\R$ and the EL operator can be conveniently written in terms of the real variable $\eta\equiv-i\xi$:
\be
\EL(x,t)=-\frac{i}{2\eta(t)}\Big[\eta^2(t)\de_x^2+\omega_0^2(1-\frac{\DE{\eta}^2(t)}{4\omega_0^2})\, x^2+\frac{1}{2}\DE{\eta}(t)\, \eta(t)(i\de_x+ix\de_x)\Big]\; ,
\ee
which can be reduced to a harmonic repulsor  (i.e. harmonic oscillator with imaginary frequency) through a unitary transformation $U=\exp(-i\DE{\eta}\, x^2/4\eta)$
\be
U\, \EL(x,t)\, U^{\dagger}=-\frac{i\eta(t)}{2}(\de_x^2+\frac{\omega^2_0\,x^2}{\eta^2(t)})\; .
\ee
From this observation, the eigenfunctions of the EL operator can be easily derived:
\be
h_{\lambda}^{\pm}(x,t)=(\frac{\omega_0}{\eta(t)})^{-1/4}\,\,  \exp(\frac{i\DE{\eta(t)}}{4\eta(t)}\, x^2)\,\, \chi_{i\lambda}^{\pm}(x\, \sqrt{\frac{\omega_0}{\eta(t)}})
\ee
where $\chi_q^{\pm}$ are the eigenfunctions of a unit frequency harmonic repulsor (see \ref{Mehler}). The spectrum of the EL operator in this case is the whole real line $\lambda \in \R$. The value of the dynamical phase $\eqref{alpha}$ in the hyperbolic case is  \cite{JER}:
\be
i\alpha(t)=-\omega_0\, \lambda \int_0^t \frac{dt'}{\eta(t')}.
\ee
In the following, given a $2\pi$-periodic solution of the equation $\eqref{Pinney}$ in terms of $\eta$ for a generic choice of the square wave frequency $\omega(t)$ in the hyperbolic domain, we derive the stroboscopic behavior of the bosonic gas. 

\subsection{Stroboscopic evolution of the one-particle wave function}
The stroboscopic evolution of the one-particle wave function in the case of hyperbolic monodromy can be deduced by plugging into the general expression $\eqref{onewave}$ the phase and eigenfunctions given above:
\be\label{onewaveH}
\psi_k(x,2\pi n)=\int_{\R} d\lambda \, \int_{\R} dy\,\, h^*_{\lambda}(y,0)\, \phi_k(y) \, e^{-nT\lambda} \, h_{\lambda}(x,0)
\ee
where we have defined 
\be
T\equiv\omega_0\int_{\gamma[0,2\pi]} \frac{dt'}{\eta(t')}
\ee
and $\gamma[0,2\pi]$ is the complex deformation of the real interval $[0,2\pi]$ which avoids the singularities, as explained in \cite{JER}. Using the definition of the hyperbolic Mehler kernel (see \ref{Mehler}) we can write $\eqref{onewaveH}$ as
\be\label{onewaveHF}
\psi_k(x,2\pi n)=\int_{\R} dy \, \exp(\frac{i\DE{\eta}_0}{4\eta_0}(x^2-y^2)) \,\, \phi_k(y) \,\, \K_{hyp}(\overline{x},\overline{y}|nT)\; ,
\ee
with the notations $ \eta_0\equiv \eta(0)$, $\DE{\eta}_0\equiv \DE{\eta}(0) $, $\overline{x}\equiv x\sqrt{\omega_0/\eta_0}$, $\overline{y}\equiv y\sqrt{\omega_0/\eta_0}$  and where the hyperbolic kernel is explicitly given by
\be\label{Khyp}
\K_{hyp}(\overline{x},\overline{y}|nT)=\sqrt{\frac{\omega_0}{\eta_0}} \,\, \frac{\exp(-\frac{i}{2}(\overline{x}^2+\overline{y}^2)\coth(nT)+i\overline{x}\,\overline{y}/\sinh(nT))}{\sqrt{2\pi\, \sinh(nT)}}\; .
\ee

\subsection{Stroboscopic evolution of the energy spectrum}
\begin{figure}
\centering
\includegraphics[scale=0.4]{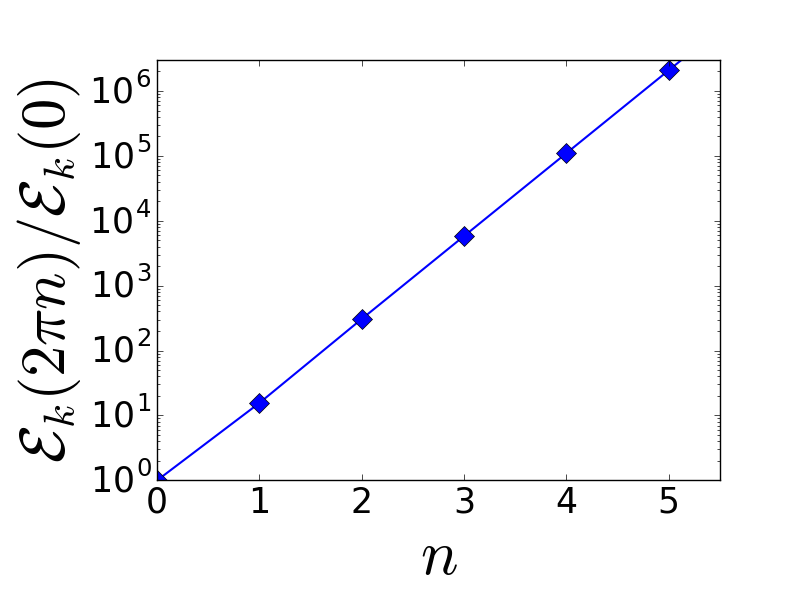}
\caption{The stroboscipic evolution of the energy spectrum $\E_k(2\pi n)/\E_k(0)$ in a logarithmic scale as function of $n\in \mathbb{N}$ (outlined points) for the hyperbolic point $(\omega_1,\omega_2)=(3/2,1/4)$ with $\tau=\pi/2$ showing an exponential growth. }\label{EH}
\end{figure}

Following the procedure used in the Sec.\ref{Eenergy}, we compute the stroboscopic evolution of the energy spectrum in the case of hyperbolic monodromy. Using the expression of the one-particle wave function $\eqref{onewaveHF}$, the computation of the expectation value $\eqref{energy}$ gives
\be\bspl
\E_{k|2}(2\pi n)=\frac{1}{2}\Big[ \Big(\cosh(nT) +\Nn \sinh(nT)\Big)^2 +\eta_0^2 \sinh^2(nT) \Big]\, \omega_0(k+\frac{1}{2})\; ,
\end{split}\ee
for the contribution of the potential and
\be\bspl
\E_{k|1}(2\pi n)=\frac{1}{2\eta_0^2}\, \Big[\Big(\cosh(nT)-\Nn\sinh(nT)\Big)^2 +\frac{1}{\eta_0^2}\sinh^2(nT)\Big(1-(\Nn)^2\Big)^2 \Big] \, \omega_0(k+\frac{1}{2})
\end{split}\ee
for the kinetic term. Hence, the result for the stroboscopic evolution of the energy spectrum is given by
\be
\E_{k}(2\pi n)=\Big[\cosh^2(nT) +(\Nn)^2\sinh^2(nT) +\frac{1}{2}\sinh^2(nT)\Big( \eta_0^2+\frac{1}{\eta_0^2}\Big(1-(\Nn)^2\Big)^2\Big) \Big] \; \omega_0(k+\frac{1}{2})\; .
\ee
Notice first that the initial value of the energy $\E_k(0)=\omega_0\, (k+1/2)$ is recovered for $n=0$.
The frequencies $\omega(t)$ in the hyperbolic monodromy class vary in time in such a way as to pump energy into the system at each cycle and the energy  grows exponentially in time. 
The physical scenario can be placed in analogy with a seesaw: depending on the behavior of the periodic forcing, the amplitude of the oscillations can either increase or remain bounded after each cycle. In Fig. \ref{EH} the stroboscopic evolution of the energy spectrum is shown for the hyperbolic point $(\omega_1,\omega_2)=(3/2,1/4)$ with $\tau=\pi/2$. 

\subsection{Large $N$ particle density stroboscopic evolution}
\begin{figure}
\centering
\includegraphics[scale=0.35]{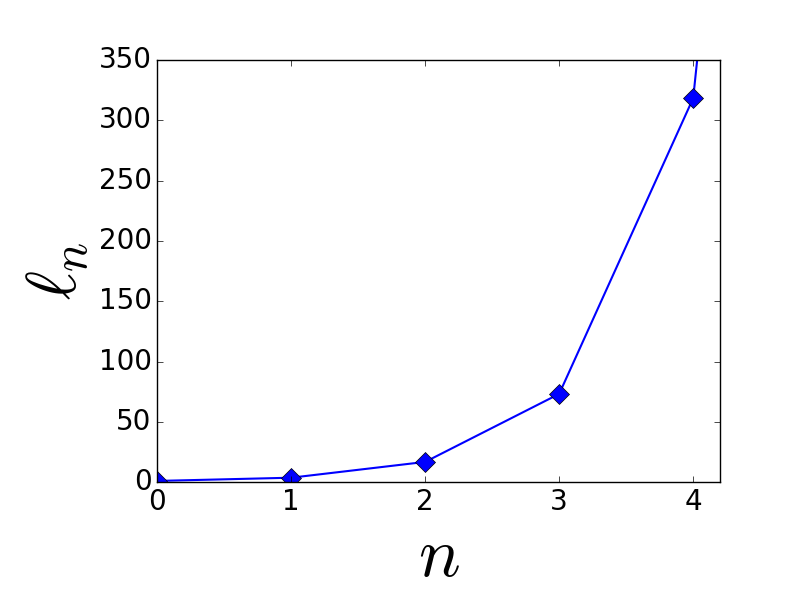}\qquad \includegraphics[scale=0.35]{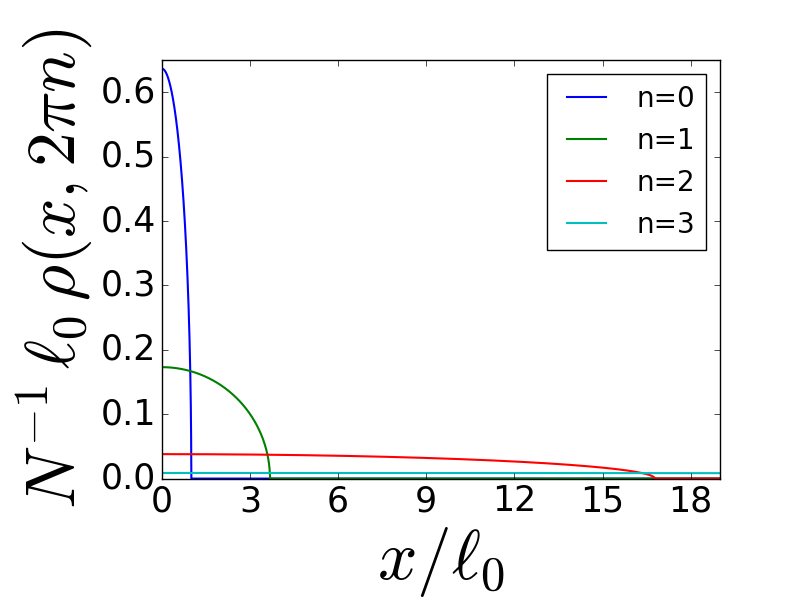}
\caption{\textsl{(Left)} The support $\ell_n$ of Eq.$\eqref{lHyp}$ for the hyperbolic point $(\omega_1,\omega_2)=(3/2,1/4)$ with $\tau=\pi/2$ as function of $n\in\mathbb{N}$. \textsl{(Right)} The associated stroboscopic evolution of the rescaled $N$-particle density $\ell_0\, \rho(x,2\pi n)/N$ $\eqref{densEH}$  as function of $\widetilde{x}=x/\ell_0$.}\label{densHYP}
\end{figure}

The stroboscopic evolution of the $N$-particle density for a system initially prepared in the ground state $\eqref{GS0}$  can be derived as in  Sec. \ref{NpartDENS} for the elliptic case. From the general expression $\eqref{generalNpart}$, using the result $\eqref{onewaveHF}$, we arrive at
\be\label{rhoNH}\bspl
\rho(x,2\pi n)= \int_{\R} dy\, \int_{\R} dy\,\,e^{i\frac{\dot{\eta}_0}{4\eta_0}(y^2-y'^2)}\;  \K_{hyp}^*(\overline{x},\overline{y}|nT) \,  \K_{hyp}(\overline{x},\overline{y}'|nT)\, \K_N(y,y') \; .
\end{split}\ee
The hyperbolic case is exactly similar to the elliptic one, up to the replacement of the kernel ${\cal K}$ by ${\cal K}_{hyp}$ and the change of notation for the phase. 
Therefore, in the limit of large number of particles $N\gg 1$, the same kind of asymptotic expansions  as in Sec.\ref{NpartDENS} leads to the final result
\be\label{densEH}
\rho(x,2\pi n) \overset{N\To\infty}{\simeq} \frac{2}{\pi}\, \frac{N}{\ell_0\, \ell_n} \, \sqrt{1-\frac{\widetilde{x}^2}{\ell^2_n}} \; ,
\ee
where the dynamical support $\ell_n$, given by 
\be\label{lHyp}
\ell_n\equiv\sqrt{\eta_0^2 \sinh^2(nT) +\Big(\cosh(nT) +\Nn \sinh(nT) \Big)^2 }\; ,
\ee
is growing exponentially in time. As a consequence,
the $N$-particle density spreads more and more in space after each period and approaches the zero-density configuration of the untrapped gas. Indeed, the external forcing gives the system more and more energy making the effect of the trap gradually negligible. The plot of the $N$-particle density and of the support $\eqref{lHyp}$ for the hyperbolic point $(\omega_1,\omega_2)=(3/2,1/4)$ with $\tau=\pi/2$ are shown in Fig. \ref{densHYP}.

%
%
%
%
%
%
%
%
%
%
%
%
%
%
%
%
%
%
\section{Summary and conclusions}
In this work we have considered a periodic forcing of a trapped one-dimensional gas of bosons with strong repulsive interactions, obtained by taking the hard core limit of a Bose-Hubbard model. 
The periodic forcing consists of a periodic variation of the harmonic trap frequency. 
We have in particular focused our attention to a square-wave type variation of the frequency, even if our general results apply to more general periodic functions.  For the square wave frequencies the classification into elliptic and  hyperbolic monodromy cases 
is very simple thanks to the correspondence of this problem with the associated Hill problem for which the monodromy matrix is obtained in a very direct way. 
The results we have obtained for the bosonic density and energy are asymptotically exact in the limit of a large number of particles, which is typically the case in a real experiment with cold atoms on an optical trap. Stroboscopically, the most significant result is that the density falls onto a semi-circle law, in both elliptic and hyperbolic cases, but with a support $\ell_n$ that varies in time periodically in the elliptic case (breathing case) and growing exponentially in time in the expanding regime (hyperbolic case).   
In the bounded case, the average density profile deviates significantly, especially at the border of the cloud, from an equilibrium semi-circle law which is a strong signature, in average, of the non-equilibrium state in which the system is.

\section*{References}

\appendix

\section{\normalfont\itshape{Mehler kernels}}\label{Mehler}
We recall the general definition of the Mehler kernel:
\be
\K(x,y|\tau)\equiv\sum_{n=0}^{\infty} \phi_n(x)\, \phi_n(y)\, e^{-(n+\frac{1}{2})\tau}=\frac{e^{-\frac{y^2}{2}+\frac{x^2}{2}}\; e^{-\frac{\tau}{2}} \, e^{-\frac{(x-ye^{-\tau})^2}{1-e^{-2\tau}}}}{\sqrt{\pi(1-e^{-2\tau})}}
\ee
where $\phi_n$ are the eigenstates of an harmonic oscillator with unit frequency. In the case of imaginary time $\tau=it$, $t\in \R$, the kernel can be written in the form:
\be
\K(x,y|it)=\frac{\exp(\frac{i}{2}(x^2+y^2)\cot(t)-i\,xy/\sin(t))}{\sqrt{2\pi i\, \sin(t)}}
\ee
and it is the Green's function of an harmonic oscillator of unit frequency, namely, $(\de_t\pm(-\de_x^2+x^2))\K=0$, see e.g. \cite{Mehler}.
The Green kernel in the hyperbolic case is by definition
\be
\K_{hyp}(x,y|\tau) = \int_{\R} dq \,\, \chi^{+\ast}_{q}(x) \, \chi^+_{q}(y)\,\, e^{i q \tau} + \,\, (+\leftrightarrow -),
\ee
where $\chi^{\pm}_{q}$ are the eigenfunctions of the unit frequency harmonic repulsor\\ $-\frac{i}{2}(\de_x^2 + x^2 )\,\chi^{\pm}_q(x)=-iq\, \chi^{\pm}_q(x)$ \cite{RogUnt}:
\be\bspl
\chi^{\pm}_{q}(x)=\frac{e^{-q \pi/4 -3i\pi/8}}{\sqrt{\pi}} \, 2^{-iq/2}\, e^{-ix^2/2} \, \Big\{ \Gamma[\frac{1}{2}(\frac{1}{2}-iq)] \, _1F_1[\frac{1}{4}-i\frac{q}{2};\frac{1}{2}; ix^2]  \\
\pm 2x\, e^{i\pi/4} \, \Gamma[\frac{1}{2}(\frac{3}{2}-iq)] \, _1F_1[\frac{3}{4}-i\frac{q}{2};\frac{3}{2}; ix^2] \Big\}\\
\end{split}\ee
in which $_1F_1$ are the Kummer's Hypergeometric functions. By construction,
\be
\int_{\R} dy \,\, \K_{hyp}(x,y|\tau) \,\, \K_{hyp}(y,z|\tau') = \K_{hyp}(x,z|\tau+\tau')\,,
\ee
i.e. $\K_{hyp}$ is a propagator,
\be\label{Khyp1}
( \de_{\tau}+ \frac{i}{2}(\de_x^2+x^2) )\, \K_{hyp}(x,y|\tau)=0,
\ee
and
\be\label{Khyp2}
\lim_{\tau\To 0} \, K_{hyp}(x,y|\tau) = \delta(x-y).
\ee
Performing a complex rotation $(x,y)\mapsto (x\, e^{i\pi/4}, y\, e^{i\pi/4})$ on the space coordinates of the real-time Mehler kernel, one sees that
\be\label{Khypdef}
\K(e^{i\pi/4}x\, ,\,  e^{i\pi/4} y\, |t) =\frac{\exp(-\frac{i}{2}(x^2+y^2)\coth(t)+i\, xy/\sinh(t))}{\sqrt{2\pi\, \sinh(t)}}
\ee
satisfies the same equation $\eqref{Khyp1}$ as $\K_{hyp}$ and the same initial condition $\eqref{Khyp2}$. Thus, the kernel $\eqref{Khypdef}$ coincides with $\K_{hyp}$.

\section{\normalfont\itshape{N-particle density}}\label{Ndensity}
We compute the wave function of $N$ hard-core bosons initially prepared in the ground state $\eqref{GS0}$ of the Hamiltonian $\eqref{Ham}$ at $t=0$:
\be
\Psi_N(\vec{x},t)=\frac{\Delta(\vec{x})}{|\Delta(\vec{x})|} \, \frac{1}{\sqrt{N!}} \, \det_{j,k=0}^{N-1} \psi_k(x_j,t)
\ee 
where $\vec{x}\equiv(x_0,\dots,x_{N-1})$ and $\Delta(\vec{x})$ is the Vandermonde determinant which symmetrize the Slater determinant under particle exchanges.\\
The $N$-particle density can be computed starting from the definition:
\be
\rho(x,t)\equiv \int_{\R^N} d\vec{x} \,\,\,  \Psi^{\ast}_N(\vec{x},t) \, \Psi_N(\vec{x},t) \,\sum_{j=0}^{N-1}\delta(x-x_j);
\ee
It is possible to write the particle density as a functional derivative of a generating functional $\mathcal{Z}[a]$:
\be\label{rho}
\rho(x,t)=\frac{\delta}{\delta a(x)}\Big\vert_{a\equiv 1} \, \mathcal{Z}[a],
\ee
where 
\be
\mathcal{Z}[a]\equiv\frac{1}{N!} \int_{\R^N} d\vec{x}\, \prod_{j=0}^{N-1} \, a(x_j) \, \det_{j,k=0}^{N-1}\psi^{\ast}_k(x_j,t) \, \det_{j',k'=0}^{N-1} \psi_{k'}(x_{j'},t).
\ee
The Andrejeff's relation \cite{andrejeff} allows us to rewrite the generating functional as:
\be\label{andrejeff}
\mathcal{Z}[a]=\det_{k,k'=0}^{N-1}\Big( \int_{\R} dx\, a(x) \, \psi^{\ast}_k(x,t) \, \psi_{k'}(x,t)\Big),
\ee
hence, the expression $\eqref{rho}$ becomes
\be\label{lastrho}
\rho(x,t)=\frac{\delta}{\delta a(x)}\Big\vert_{a\equiv 1} \, \mathcal{Z}[a]= \det_{k,k'=0}^{N-1} B_{k,k'}[1] \, \cdot \,  \Tr\Big[ (B_{k,k'}[1])^{-1} \, \frac{\delta B_{k,k'}[a]}{\delta a(x)}\Big\vert_{a\equiv 1}\Big],
\ee
where we have introduced the definition
\be
B_{k,k'}[a]\equiv \int_{\R} dx \, a(x)\,  \psi^{\ast}_k(x,t) \, \psi_{k'}(x,t).
\ee
Using the result $\eqref{onewave}$ for the one-particle wave function we can compute the matrix elements $B_{k,k'}[a]$ explicitly. The results are
\be\bspl
B_{k,k'}[1]=\int_{\R} dx \,  \psi^{\ast}_k(x,t) \, \psi_{k'}(x,t)= \delta_{k,k'}.
\end{split}
\ee
and
\be
\frac{\delta B_{k,k'}[a]}{\delta a(x)} \Big\vert_{a\equiv 1} = \psi_{k}^{\ast}(x,t) \, \psi_{k'}(x,t).
\ee
Inserting the last two results into $\eqref{lastrho}$ we obtain
\be
\rho(x,t)=\sum_{k=0}^{N-1} |\psi_k(x,t)|^2
\ee
which is a well-known result \cite{Scopa,MING,VIC}.

\section{\normalfont\itshape{Monodromy matrix for a square-wave frequency}}\label{piece-mon}
We investigate the monodromy of an harmonic Hamiltonian $\eqref{Ham}$ with square-ware frequency $\eqref{piece}$ considering the associated Hill's equation:
\be
\DEE{y}(t)+\omega(t)\, y(t)=0\,,
\ee
with initial conditions $y(0)$ and $\DE{y}(0)$. In the interval $0\leq t\leq \tau$, where $\omega(t)=\omega_1$, a set of independent solutions is given by 
\be
Y(t)\equiv \bV y(t) \\ \DE{y}(t)\eV= \bV \cos(\omega_1t) & \sin(\omega_1 t)/\omega_1 \\ -\omega_1\sin(\omega_1 t) & \cos(\omega_1 t) \eV \, \bV y(0) \\ \DE{y}(0) \eV .
\ee
For times $\tau\leq t \leq 2\pi$, we can proceed in the same way updating the initial conditions to $Y(\tau)$. The solutions now reads
\be
Y(t)=\bV \cos(\omega_2(t-\tau)) & \sin(\omega_2 (t-\tau))/\omega_2 \\ -\omega_2\sin(\omega_2 (t-\tau)) & \cos(\omega_2 (t-\tau)) \eV \, Y(\tau).
\ee
From the definition $\eqref{mon-matrix}$ and using the last results we obtain:
\be
\mathbb{M}(\omega)=\bV \cos(\omega_2(2\pi-\tau)) & \sin(\omega_2 (2\pi-\tau))/\omega_2 \\ -\omega_2\sin(\omega_2 (2\pi-\tau)) & \cos(\omega_2 (2\pi-\tau)) \eV\bV \cos(\omega_1t) & \sin(\omega_1 \tau)/\omega_1 \\ -\omega_1\sin(\omega_1 \tau) & \cos(\omega_1 \tau) \eV 
\ee
with trace given by \cite{Magnus-Winkler}\footnote{Notice that a factor $1/2$ is missing in the second term of this expression in \cite{Magnus-Winkler}. }
\be
\text{Tr}(\mathbb{M}(\omega))=2\cos(\omega_1\tau)\cos(\omega_2(2\pi-\tau)) -\Big(\frac{\omega_1}{\omega_2}+\frac{\omega_2}{\omega_1}\Big) \sin(\omega_1\tau)\sin(\omega_2(2\pi-\tau)).
\ee

\qqq